\documentstyle[aps,twocolumn,amssymb]{revtex}   

\begin{document}

\draft
\preprint{UCR--FM--03--94}

\title{From geometric quantization to Moyal quantization}
\author{Jos\'e M. Gracia-Bond{\'\i}a}
\address{Departamento de F{\'\i}sica Te\'orica,
         Universidad Aut\'onoma de Madrid, 28049 Madrid, Spain \\
         {\rm and} \\
         Departamento de F{\'\i}sica Te\'orica,
         Universidad de Zaragoza, 50009 Zaragoza, Spain}
\author{Joseph C. V\'arilly}
\address{Escuela de Matem\'atica, Universidad de Costa Rica,
         San Jos\'e, Costa Rica}
\date{\today}
\maketitle
\begin{abstract}
We show how the Moyal product of phase-space functions, and the Weyl
correspondence between symbols and operator kernels, may be obtained
directly using the procedures of geometric quantization, applied to the
symplectic groupoid constructed by ``doubling'' the phase space.
\end{abstract}
\pacs{PACS numbers: 03.65.Fd, 02.40.Ma.}

\narrowtext

\section{Introduction}

Over the last two decades, several approaches to quantization of
classical systems have been developed. The most
mathematically thorough of these is the so-called method of geometric
quantization \cite{Kostant,Kirillov,Robinson,Woodhouse}, which seeks to
manufacture the quantum-mechanical Hilbert space from the symplectic
manifold of classical states. Other quantization procedures may be
refined and extended by recasting them in the geometric quantization
framework; thus, for example, the recent work of Tuynman \cite{Tuynman}
on BRST symmetry. The relations between different quantization schemes
continue to merit attention \cite{Bordemann,Borthwick}.

   The Moyal or phase-space approach to quantization
\cite{Moyal,Bayenetal,Iapetus} has not, so far, been explicitly derived
from the geometric quantization scheme. This was pointed out by
Weinstein \cite{WeinsteinNCG}. However, we are not aware of an explicit
treatment in the literature; this note attempts to fill that gap. We
spell out how these two approaches may be related, in the simplest case
of a linear phase space. The idea needed to bridge the gap between both
quantization schemes is the concept of symplectic groupoid, developed by
Weinstein and co-workers
\cite{WeinsteinBull,WeinsteinNCG,WeinsteinTorus,WeinsteinXu}.

   The article is arranged as follows. In Sec.~II we recall the
definition of symplectic groupoid, and in Sec.~III we briefly review
the theory of pairings in geometric quantization, in order to
establish the context. In Sec.~IV we show that the Weyl correspondence
between Weyl symbols of Hilbert--Schmidt operators on $L^2({\Bbb R}^n)$
and their kernels, is given by a pairing of real polarizations of a
particular symplectic groupoid, namely two copies of the flat
phase-space ${\Bbb R}^{2n}$. We then show, in Sec~V, that the Moyal
product of phase-space functions arises directly from the groupoid
structure of the double phase space.

   Two further applications of this viewpoint are given. In Sec.~VI, we
rederive the integral transformation introduced by Daubechies and
Grossmann \cite{Daubechies} to effect quantization in the
coherent-state picture, from a pairing of a real and a complex
polarization on the aforementioned groupoid. Finally, it is shown in
Sec.~VII that the appearance of the ordinary Fourier transformation as a
power of the Weyl correspondence map can be understood geometrically as
a  property of symplectic transformations on that groupoid.

\section{Symplectic groupoids}

If $M$ is a manifold with symplectic form~$\omega$, we will denote by
$\overline M$ the symplectic manifold $(M,-\omega)$. A groupoid is a set
with a partially-defined associative multiplication. We recall the
definition of a symplectic groupoid, as  set forth
in~\cite{WeinsteinNCG}.

   A {\it symplectic groupoid\/} consists of a pair of mani\-folds
$(G,G_0)$, where $G$ has a symplectic form $\Omega$ and a
partially defined multiplication with domain
$G_2 \subset G \times G$, together with two submersions
$\alpha\colon G \to G_0$, $\beta\colon G \to G_0$,
and an involution $x \mapsto x^*$ of $G$, such that:

\begin{enumerate}

\item
the graph ${\cal M} = \{ (x,y,xy) : (x,y) \in G_2 \}$ of the
multiplication is a Lagrangian submanifold of
$\overline G \times \overline G \times G$;

\item
the set of ``units'' $G_0$ may be identified with a Lagrangian
submanifold of $G$ (also denoted by $G_0$);

\item
for any $x \in G$, we have $\alpha(x)x = x = x\beta(x)$; and
$\alpha(x) = xx^*$, $\beta(x) = x^*x$; moreover, $(x,y) \in G_2$ iff
$\beta(x) = \alpha(y)$;

\item
the graph $I = \{ (x,x^*) : x \in G \}$ of the involution is
a Lagrangian submanifold of $G \times G$;

\item
whenever $(x,y)$ and $(y,z) \in G_2$, then $(xy,z)$ and $(x,yz)$ lie in
$G_2$, and $(xy)z = x(yz)$.

\end{enumerate}

   As consequences of these postulates, we find that
$\alpha(x^*) = \beta(x)$; that $\alpha(x)^* = \alpha(x) = \alpha(x)^2$
and $\beta(x)^* = \beta(x) = \beta(x)^2$; that $xx^*x = \alpha(x)x = x$;
that $\alpha(\alpha(x)) = \alpha(x)$ and $\beta(\beta(x)) = \beta(x)$.
Moreover, if $(x,y) \in G_2$, then
$$
\alpha(xy) = xyy^*x^* = x\alpha(y)x^* = x\beta(x)x^* = xx^* = \alpha(x),
$$
and also $\beta(xy) = \alpha(y^*x^*) = \alpha(y^*) = \beta(y)$.

   As a notational convention, we write $G \rightrightarrows G_0$ to
denote a symplectic groupoid, if $\alpha$ and $\beta$ are understood.

   Two general examples of symplectic groupoids deserve mention. One is
the groupoid $T^*H \rightrightarrows {\frak h}^*$, where $H$ is a Lie
group and ${\frak h}^*$ is the dual of its Lie algebra. The maps
$\alpha$ and $\beta$ are given by right, resp.\ left, translation of a
cotangent vector to the cotangent space at the identity of~$H$.

   Another example is the fundamental groupoid
$\pi(M) \rightrightarrows M$ of a symplectic manifold $(M,\omega)$.
Its elements are homotopy classes of smooth paths
$\sigma\colon [0,1] \to M$, with the usual concatenation product of
paths whose endpoints match; reversing the path gives the involution.
Here $\alpha([\sigma]) = \sigma(0)$,
$\beta([\sigma]) = \sigma(1)$ are the endpoint assignment
maps. The manifold $M$ embeds in $\pi(M)$ as the submanifold of
constant paths, which is Lagrangian with respect to the symplectic
structure $\Omega = \alpha^*\omega - \beta^*\omega$
on~$\pi(M)$.

   When $M$ is simply connected, $[\sigma]$ is determined by its
endpoints, and the fundamental groupoid may be reexpressed as
\[
M \times \overline M \rightrightarrows M.
\]
We can then write $\alpha(q,p) = q$,
$\beta(q,p) = p$, and identify $M$ with the diagonal
submanifold $\{ (q,q) : q \in M \}$. The multiplication and
involution are given by:
\[
(q,p) \cdot (p,r) = (q,r);  \quad  (q,p)^* = (p,q).
\]
One checks that the graph of the product
${\cal M} = \{ (q,p;p,r;q,r) : q,p,r \in M \}$ is Lagrangian
in $\overline G \times \overline G \times G$.

   We now specialize further to the case $M = {\Bbb R}^{2n}$, with
$\omega$ a nondegenerate alternating bilinear form on~${\Bbb R}^{2n}$.
Writing $\hat\omega(u) \colon v \mapsto \omega(u,v)$ gives a
skewsymmetric invertible map
$\hat\omega \colon {\Bbb R}^{2n} \to {\Bbb R}^{2n*}$. One obtains
\begin{eqnarray*}
&&\Omega((x,y),(z,w)) = \omega(x,z) - \omega(y,w)  \\
&&\quad = \hat\omega(x - y) \Bigl[{z + w \over 2}\Bigr]
        - \hat\omega(z - w) \Bigl[{x + y \over 2}\Bigr].
\end{eqnarray*}

   On the other hand, ${\Bbb R}^{2n} \times \overline{{\Bbb R}}^{2n}$
can be identified with the cotangent bundle $T^*({\Bbb R}^{2n})$. If
$(u,\xi)$, $(v,\eta)$ are elements of
${\Bbb R}^{2n} \times {\Bbb R}^{2n*}$, regarded as local coordinates
of covectors in $T^*({\Bbb R}^{2n})$, the cotangent symplectic structure
of $T^*({\Bbb R}^{2n})$ reduces to the alternating bilinear form:
\[
\Sigma((u,\psi), (v,\chi)) = \chi(u) - \psi(v).
\]
Thus ${\Bbb R}^{2n} \times \overline{{\Bbb R}}^{2n}$ can be identified
with $T^*({\Bbb R}^{2n})$ as symplectic manifolds by the linear
isomorphism
\begin{equation}
\Phi\colon (x,y) \mapsto \bigl( \case12(x+y), \hat\omega(x-y) \bigr)
\label{Phidefn}
\end{equation}
for which $\Phi^*\Sigma = \Omega$.

\section{Pairing in geometric quantization}

We briefly recall here, in order to fix notation, those aspects of
geometric quantization we will need to address.

   Prequantization of an $2n$-dimensional symplectic manifold
$(M,\omega)$ proceeds by finding a real-linear map $f \mapsto \hat f$
from the Poisson algebra of smooth functions on~$M$ to an algebra of
operators on the Hilbert space of $L^2(M)$, for which $\hat 1 = I$ and
$\{f_1,f_2\}\,\widehat{}{} = (i/\hslash)[\hat f_1,\hat f_2]$. The right
recipe is $\hat f = f - i\hslash\,\nabla_{X_f}$, where $X_f$
is the Hamiltonian vector field of~$f$ and the covariant derivative
$\nabla$ is locally given by
\begin{equation}
\nabla_X = X - (i/\hslash)\,\theta(X).
\label{nabla}
\end{equation}
Here $\theta$ is a symplectic potential, i.e., a one-form for which
$d\theta = \omega$. When $\omega$ is not exact, local potentials must
be patched together so that $\nabla$ becomes a linear connection on a
Hermitian complex line bundle $L \to M$, whose curvature form is
$(-i/\hslash)\omega$, as is well-known. The elements of the
prequantization Hilbert space are sections $s \in \Gamma L$ of this line
bundle.

   Geometric quantization then involves finding a positive
polarization of $(M,\omega)$, i.e., a subbundle $F$ of the complexified
tangent bundle $T^*M^{\Bbb C}$, which is maximally isotropic
for~$\omega$, with $F \cap \overline F$ of constant rank; which is
integrable in the sense that both $F$ and $F \cap \overline F$ are
closed under the Lie bracket; and which is positive in that
$-i\omega(\bar Y, Y) \geq 0$ whenever $Y$ is a section of~$F$.

   A polarized section is any $s \in \Gamma L$ for which
$\nabla_Y s = 0$ whenever $Y \in \Gamma F$. The quantizable observables
are those $g \in C^\infty(M)$ for which $\mathop{\rm ad}(X_g)$
preserves $\Gamma F$. Then one checks that $\hat g$ preserves the space
$\Gamma_F L$ of polarized sections. The remaining difficulty is to
endow $\Gamma_F L$ ---or some modification thereof--- with a suitable
inner product, in order that the quantizable observables be represented
as operators on a Hilbert space. This is done by using the idea of a
half-form {\it pairing\/}~\cite{Blattner}.

   We follow the very precise treatment of pairings by Rawnsley
\cite{RawnsleyPair,RawnsleyKepler}. The canonical line bundle of $F$
is $K^F = \Lambda^n F^0$, where $F^0 \subset T^*M^{\Bbb C}$
denotes covectors which vanish on~$F$. For example, if $M$ is a
K\"ahler manifold with local holomorphic coordinates $(z_1,\dots,z_n)$,
and $F$ is spanned by
$\partial/\partial\bar z_1,\dots,\partial/\partial\bar z_n$, then $K^F$
is spanned by $dz_1 \wedge\cdots\wedge dz_n$; in this case we have
$F \cap \overline F = 0$. A contrasting example, for which $F$ is a
real polarization, that is, $F = \overline F$, is obtained by taking
local Darboux coordinates $(q_1,\dots,q_n,p_1,\dots,p_n)$ for~$M$, with
$F$ spanned by $\partial/\partial p_1,\dots,\partial/\partial p_n$,
whereupon $K^F$ is spanned by $dq_1 \wedge\cdots\wedge dq_n$.

   Suppose we have two positive polarizations $F$ and $P$; it turns out
that $K^F$ and $K^P$ are isomorphic as line bundles over~$M$ and that
$\overline{K^F} \otimes K^P$ is a trivial bundle. There is an obvious
map from this bundle to $\Lambda^{2n} T^*M^{\Bbb C}$ (replace tensor by
exterior product), which is an isomorphism iff $\overline F \cap P = 0$.
The Liouville volume $\lambda = (-1)^{n(n-1)/2} \omega^{\wedge n}/n!$
trivializes the latter bundle. Thus we have a pairing
$\langle \alpha,\beta \rangle$ of $\alpha \in \Gamma K^F$ and
$\beta \in \Gamma K^P$ defined by
\begin{equation}
i^n \langle \alpha,\beta \rangle\, \lambda
 =  \bar\alpha \wedge \beta
\label{pairingdefn}
\end{equation}
provided $\overline F \cap P = 0$. In particular, if
$\overline F \cap F = 0$, then $\langle\cdot,\cdot\rangle$ is an inner
product on $\Gamma K^F$.

   Matters are less straightforward if $\overline F \cap P \neq 0$.
Here $\overline F \cap P = D^{\Bbb C}$ where $D$ is an isotropic
subbundle of~$TM$. If $D^\perp$ is the symplectic orthogonal of~$D$,
then $D^\perp/D$ becomes a symplectic vector bundle (with an induced
symplectic form $\omega_D$), of which $\overline F/D$ and $P/D$ are
nonoverlapping maximal-isotropic subbundles; thus we may apply the
previous recipe to get a pairing of $K^{F/D}$ and $K^{P/D}$.

   We can try to pull back to a pairing of $K^F$ and $K^P$ by
suppressing the common real directions in~$D$. Suppose that the
foliation of~$M$ induced by~$D$ has a smooth space of leaves $M/D$,
that $D$ is spanned locally by
$\partial/\partial y_1,\dots,\partial/\partial y_k$, and that
$(x_1,\dots,x_k)$ are conjugate local coordinates to $(y_1,\dots,y_k)$;
if $\alpha = a \,dx_1 \wedge\cdots\wedge dx_k
  \wedge dz_1 \wedge\cdots\wedge dz_{n-k} \in \Gamma K^F$,
$\beta = b \,dx_1 \wedge\cdots\wedge dx_k
  \wedge dw_1 \wedge\cdots\wedge dw_{n-k} \in \Gamma K^P$, where the
coefficient functions do not depend on the $y_j$, then we can define
$\tilde\alpha =
  a \,d\tilde z_1 \wedge\cdots\wedge d\tilde z_{n-k}$ in
$\Gamma K^{F/D}$, $\tilde\beta =
  b \,d\tilde w_1 \wedge\cdots\wedge d\tilde w_{n-k}$ in
$\Gamma K^{P/D}$, where the tildes denote corresponding coordinates on
$M/D$, and we can try to set $\langle\alpha,\beta\rangle  =
  \langle\tilde\alpha,\tilde\beta\rangle$. It turns out, of course,
that this recipe is coordinate-dependent, and in fact (after
incorporating a correction factor of~$\lambda^2$) the change of
variables formula shows that the result is a 2-density on the leaf
space $M/D$.

   Since we could integrate a 1-density over $M/D$ to get a
scalar-valued inner product, we abandon $K^F$ in favour of the vector
bundle $Q^F$ of ``half-forms'' on~$M$ which is defined by the
requirement that $Q^F \otimes Q^F = K^F$; if $\alpha \in \Gamma K^F$,
we write $\sqrt\alpha =  \mu \in \Gamma Q^F$ if
$\mu \otimes \mu = \alpha$. It can then be shown that
$\overline{Q^F} \otimes Q^P$ carries a pairing, whose values are
1-densities on~$M/D$, determined (up to a sign) by the requirement that
$\langle\sqrt\alpha, \sqrt\beta\rangle^2 = \langle\alpha,\beta\rangle$.

   (We tiptoe past the crucial question of the existence of $Q^F$, for
which there is a topological obstruction: $(M,\omega)$ must ``admit
metaplectic structures''. This obstruction has been ingeniously
overcome by Robinson and Rawnsley \cite{Robinson} by replacing
metaplectic structures by $Mp^c$-structures, which always exist; the
procedure is akin to passing from spin structures to spin$^c$
structures on Riemannian manifolds.)

   The final touch is to replace the prequantization bundle $L$ by
$L \otimes Q^F$, and let $\Gamma_F(L \otimes Q^F)$ denote its
polarized sections (those killed by $\nabla_Y$ for $Y \in \Gamma F$).
The pairing of two sections
$s \otimes \sqrt\alpha \in \Gamma_F(L \otimes Q^F)$,
$t \otimes \sqrt\beta \in \Gamma_P(L \otimes Q^P)$ is given by
\begin{equation}
\langle s \otimes \sqrt\alpha, t \otimes \sqrt\beta \rangle =
\int_{M/D} (s,t) \, \langle \sqrt\alpha, \sqrt\beta \rangle,
\label{fullpairing}
\end{equation}
where $(\cdot,\cdot)$ is the Hermitian metric on~$L$. When $F = P$, the
geometric quantization Hilbert space ${\cal H}_F$ is obtained by
completing $\Gamma_F(L \otimes Q^F)$ with respect to this inner product.

\section{Pairings and the Weyl correspondence}

On the symplectic manifold $G_0 = {\Bbb R}^{2n}$, we take coordinates
$(x',x'') \equiv (x'_1,\dots,x'_n,x''_1,\dots,x''_n)$, so that
$\omega = dx' \wedge dx'' \equiv \sum_k dx'_k \wedge dx''_k$. (To avoid
index clutter, we will henceforth just take $n = 1$.) We can regard
$\omega$ as a bilinear symplectic form on ${\Bbb R}^2$, with
$\omega(x,z) = x'z'' - x''z'$. Then
$\hat\omega(x) = (-x'',x')$ in the dual space ${\Bbb R}^{2*}$.

   The symplectic groupoid $G = {\Bbb R}^2 \times \overline{\Bbb R}^2$
has coordinates $(x',x'';y',y'')$, with which its symplectic form may
be written as
\begin{equation}
\Omega = \pi_1^*\omega - \pi_2^*\omega
 = dx' \wedge dx'' - dy' \wedge dy''.
\label{Omeg}
\end{equation}
Thus $(x',y';x'',-y'')$ are Darboux coordinates for~$G$.

   On the cotangent bundle $T^*{\Bbb R}^2$, we use Darboux coordinates
$(q_1,q_2,p_1,p_2)$; the symplectic form
$\Sigma = dq_1 \wedge dp_1 + dq_2 \wedge dp_2$. The symplectomorphism
$\Phi$ of (\ref{Phidefn}) is given explicitly by
\begin{eqnarray}
&& q_1 = {x' + y' \over 2},\ q_2 = x' - y', \nonumber\\
&& p_1 = x'' - y'',\         p_2 = {x'' + y'' \over 2}.
\label{qpxy}
\end{eqnarray}

   We consider the following two real polarizations of $G$. Set
\begin{eqnarray}
F &=& \text{span} \biggl\{ {\partial \over \partial x''},
                  {\partial \over \partial y''} \biggr\}, \nonumber \\
P &=& \text{span} \biggl\{ {\partial \over \partial p_1},
                  {\partial \over \partial q_2} \biggr\}.
\label{polzns}
\end{eqnarray}
{}From (\ref{qpxy}), we have
\[
{\partial \over \partial p_1}
 = {1\over2} \biggl( {\partial \over \partial x''}
                   - {\partial \over \partial y''} \biggr), \quad
{\partial \over \partial p_2}
 = {\partial \over \partial x''} + {\partial \over \partial y''},
\]
so we can rewrite
$F = \text{span}\{ \partial/\partial p_1, \partial/\partial p_2 \}$.
Therefore $\overline F \cap P = D^{\Bbb C}$ where $D$ is spanned by
$\partial/\partial p_1$. By a slight abuse of notation, we can regard
$\{ q_1, q_2, p_2 \}$ as local coordinates for the (affine) leaf
space $G/D$, and the pairing
$\Gamma Q^F \times \Gamma Q^F \to {\cal D}^1(G/D)$ is determined by
\[
\langle \sqrt{dx' \wedge dy'}, \sqrt{dq_1 \wedge dp_2} \rangle
 = dq_1 \,dq_2 \,dp_2.
\]

   The polarized sections in $\Gamma_F L$ are of the form $fs_0$, where
$f \in C^\infty(G)$ and $s_0$ is a nonvanishing section of~$L$
satisfying $\nabla_X s_0 = - (i/\hslash)\,\Theta_F(X)s_0$ and
$(s_0, s_0) = 1$. The symplectic potential $\Theta_F$ for $(G,\Omega)$
may be taken to vanish on~$F$; and so
\[
\Theta_F = - x''\,dx' + y''\,dy' = - p_1\,dq_1 - p_2\,dq_2.
\]
In this case $fs_0 \in \Gamma_F L$ iff $Xf = 0$ for $X \in F$, that is,
$f = f(x',y')$. Likewise, if $t_0$ is a section of~$L$ satisfying
$\nabla_X t_0 = - (i/\hslash)\,\Theta_P(X)t_0$ and $(t_0, t_0) = 1$,
with
\[
\Theta_P = - p_1\,dq_1 + q_2\,dp_2
\]
being the symplectic potential which vanishes on~$P$, then a typical
element of $\Gamma_P L$ is of the form $gt_0$ with $g = g(q_1,p_2)$.

   Clearly $t_0 = \phi_0s_0$ for a nonvanishing
$\phi_0 \in C^\infty(G)$; indeed, from
$\nabla_X t_0 = (X\phi_0)s_0 + \phi_0\,\nabla_X s_0$ we obtain
\[
{d\phi_0 \over \phi_0} = {i\over\hslash} (\Theta_F - \Theta_P)
 = -{i\over\hslash} d(q_2p_2),
\]
and so $\phi_0 = C\, \exp(-iq_2p_2/\hslash)$ for some positive
constant~$C$. Since $(s_0, t_0) = \phi_0$, we can now compute the
half-form pairing of
$\alpha = f(x',y') s_0 \otimes \sqrt{dx' \wedge dy'}$ and
$\beta = g(q_1,q_2) t_0 \otimes \sqrt{dq_1 \wedge dp_2}$ as
\begin{eqnarray*}
&&\langle\alpha, \beta\rangle
  = C \int \overline{f(x',y')} g(q_1,p_2) \,e^{-iq_2p_2/\hslash}
            \,dq_1 \,dq_2 \,dp_2  \\
&&= C \int \overline{f(x',y')} g\Bigl( {x'+y'\over2}, p_2 \Bigr)
            \,e^{ip_2(y' - x')/\hslash} \,dp_2 \,dx' \,dy'  \\
&&= \langle f, Tg \rangle_{L^2({\Bbb R}^2)},
\end{eqnarray*}
where
\begin{equation}
Tg(x',y') := C \int g\Bigl( {x'+y'\over2}, \zeta \Bigr)
                     \,e^{i\zeta(y' - x')/\hslash} \,d\zeta
\label{Wigner-tr}
\end{equation}
is the kernel of the operator ---on $L^2({\Bbb R})$--- whose Weyl
symbol is~$g$ \cite{Hormander}. Unitarity of~$T$ is achieved by taking
$C = (2\pi\hslash)^{-1}$.

   In other words: the pairing of the non-transverse polarizations $F$
and~$P$ of the symplectic groupoid
${\Bbb R}^2 \times \overline{\Bbb R}^2$ yields the well-known
correspondence between kernels of Hilbert--Schmidt operators on
$L^2({\Bbb R})$ and the Weyl symbols of these operators. Thus the
groupoid forms a bridge between conventional quantum mechanics and
the phase-space formalism. It remains only to see how the symbol
product may be obtained directly from this starting point.

\section{The Moyal product from geometric quantization}

The importance of symplectic groupoids in general is that the partial
multiplication in~$G$ induces an associative product of polarized
sections, so that {\it the geometric quantization Hilbert space becomes
in fact a Hilbert algebra}. By suitably modifying its topology, one can
obtain a $C^*$-algebra. This is in the spirit of noncommutative geometry
\cite{ConnesIHES,ConnesBook,Sirius}. Indeed, in
\cite{WeinsteinTorus,WeinsteinTheta}, a symplectic groupoid structure on
the torus ${\Bbb T}^2$, which depends on an irrational parameter, is
shown to yield the ``noncommutative torus'' algebra considered by
Rieffel and others \cite{ConnesBook,Rieffel}.

   On the other hand, the basic idea of Moyal quantization is that by
working with functions on phase space, rather than wave functions, one
may describe both states and observables of quantum-mechanical systems
in classical terms; thus phase-space functions are to be equipped
with a noncommutative product which give the quantum formalism directly
without invoking a Hilbert space a priori. In
Ref.~\onlinecite{WeinsteinNCG} it is claimed that the Moyal product of
phase-space functions is inherited from the groupoid structure of
${\Bbb R}^2 \times \overline{\Bbb R}^2 \rightrightarrows {\Bbb R}^2$,
equipped with the polarization $P$ of~(\ref{polzns}). We next verify
this claim in detail.

   For any groupoid~$G$, we may define a convolution product of two
functions $f$, $g$ on~$G$ by
\[
(f * g)(z) := \int_{\{xy=z\}} f(x) g(y) \,d\lambda_z(x,y),
\]
where $\lambda_z$ is some suitable measure on the set
$\{ (x,y) \in G_2 : xy = z \}$. For the symplectic groupoid
$G = M \times \overline M$, this simplifies to:
$$
(f * g)(x,y) := \int_M f(x,t) g(t,y) \,d\lambda(t),
$$
where $\lambda = \lambda_{x,y}$ is (a multiple of) the Liouville volume
on~$M$.

   When $G$ has a real polarization with a regular leaf space, the
polarized sections are represented (locally) by functions covariantly
constant along the leaves; in general their convolution products will
fail to be covariantly constant. To obtain a new polarized section,
one must average over the leaves (by integration); by projection, one
recovers a {\it twisted product\/} of functions on the leaf space.

   In the case $G = {\Bbb R}^2 \times \overline{\Bbb R}^2$, the diagonal
$\Delta = \{ (x',x'';x',x'') \in G : (x',x'') \in {\Bbb R}^2 \}$ is
a Lagrangian submanifold of~$G$ which is transverse to the leaves
$q_1 = \text{const}_1$, $q_2 = \text{const}_2$ of the polarization $P$;
thus a polarized section is determined by its values on~$\Delta$, and
we may identify $\Delta$ with the leaf space $G/P$.

   Let us now regard Eq.~(\ref{qpxy}) as a linear change of variables;
we wish to rewrite the groupoid product
\begin{equation}
(x',x'',y',y'') = (x',x'',t',t'') \cdot (t',t'',y',y'')
\label{old-prod}
\end{equation}
in a more suitable form; we substitute
\begin{eqnarray}
&&q = \case12 (x' + y'),\ q' = \case12 (x' + t'),\
  q'' = \case12 (t' + y');  \nonumber \\
&&p = \case12 (x'' + y''),\ p' = \case12 (x'' + t''),\
  p'' = \case12 (t'' + y'');  \nonumber \\
&&\xi = x'' - y'',\ \xi' = x'' - t'',\ \xi'' = t'' - y''; \nonumber \\
&&\eta = y' - x',\ \eta' = t' - x',\ \eta'' = y' - t'.
\label{new-coords}
\end{eqnarray}
Now Eq.~(\ref{old-prod}) takes the form
\begin{equation}
(q,p,\xi,\eta) = (q',p',\xi',\eta') \cdot (q'',p'',\xi'',\eta''),
\label{new-prod}
\end{equation}
determined by the four relations
\begin{eqnarray}
q    &=& \case12 (q' + q'') - \case14 (\eta' - \eta''),  \nonumber \\
p    &=& \case12 (p' + p'') + \case14 (\xi' - \xi''),  \nonumber \\
\xi  &=& 2(p' - p''),  \nonumber \\
\eta &=& 2(q'' - q').
\label{prod-rels}
\end{eqnarray}
Now $\alpha(q,p,\xi,\eta) = (q - \case12 \eta, p + \case12 \xi)$,
$\beta(q,p,\xi,\eta) = (q + \case12 \eta, p - \case12 \xi)$ in the new
coordinates, so the partial product (\ref{new-prod}) is subject to the
compatibility conditions:
\begin{eqnarray}
q' + \case12 \eta' &=& q'' - \case12 \eta'',  \nonumber \\
p' - \case12 \xi'  &=& p'' + \case12 \xi''.
\label{compat}
\end{eqnarray}
We may interpret the coordinate change (\ref{new-coords}) thus: the
parameters $(q,p)$ label points of the leaf space $G/P$ (since
$\Delta$ is the submanifold $\xi = \eta = 0$), while $(\xi,\eta)$ are
parameters along the leaves. Since $(x',x'',y',y'')
 = (q - \case12\eta, p + \case12\xi, q + \case12\eta, p - \case12\xi)$,
each leaf carries a natural volume form $2^{-4}\,d\eta \wedge d\xi$.

   The pointwise product of two functions on~$G$ representing sections
in $\Gamma_P(L \otimes Q^P)$ is
\[
(2\pi\hslash)^{-2} g(q',p') e^{-ip'\eta'/\hslash} \,
 h(q'',p'') e^{-ip''\eta''/\hslash},
\]
which is of the form
\[
f(q,p,q',p',q'',p'') e^{-ip\eta/\hslash}
\]
with
\begin{eqnarray}
&& f(q,p,q',p',q'',p'')  \nonumber  \\
&& = (2\pi\hslash)^{-2} g(q',p') h(q'',p'')
   \exp\Bigl( -{i\over\hslash} (p'\eta' + + p''\eta'' - p\eta) \Bigr)
   \nonumber  \\
&& = (2\pi\hslash)^{-2} g(q',p') h(q'',p'')
\label{exp-mess}  \\
&&\qquad \times  \exp\Bigl(-{2i\over\hslash}
      (pq' - qp' + p'q'' - q'p'' + p''q - q''p) \Bigr), \nonumber
\end{eqnarray}
since the relations Eqs.~(\ref{prod-rels}) and (\ref{compat}) imply
\[
\eta = 2(q'' - q'),\ \eta' = 2(q'' - q),\ \eta'' = 2(q - q').
\]

   The twisted product $(g \times h)(q,p)$ is thus an integral of the
expression (\ref{exp-mess}) over: (a) the parameter region
$(t',t'') \in {\Bbb R}^2$ determined by (\ref{compat}) which underlies
the (prequantized) convolution product, and (b) the leaf of $P$
through the point $(q,p) \in \Delta$, which is parametrized by
$(q - \case12 \eta, p + \case12 \xi)$. Since
\begin{eqnarray*}
&& dt' \wedge dt'' \wedge (2^{-4}\, d\eta \wedge d\xi)  \\
&& = \case 14\, d(q' + q'') \wedge d(p' + p'')
         \wedge d(q'' - q') \wedge d(p' - p'')  \\
&& = dq' \wedge dq'' \wedge dp' \wedge dp'',
\end{eqnarray*}
we finally arrive at
\begin{eqnarray*}
&& (g \times h)(q,p)
   = (2\pi\hslash)^{-2} \int_{{\Bbb R}^4} g(q',p') h(q'',p'')  \\
&&\qquad \times \exp\Bigl(-{2i\over\hslash}
   (pq' - qp' + p'q'' - q'p'' + p''q - q''p) \Bigr)  \\
&&\qquad \times  \,dq'\,dq''\,dp'\,dp'',
\end{eqnarray*}
which is the Moyal product \cite{Moyal,Iapetus} of the symbols $g$
and~$h$. Thus the geometric quantization data $(G,\Omega,P)$ indeed
incorporate the essentials of Moyal quantization in the linear case.

\section{The Daubechies--Grossmann transform}

Some years ago, Daubechies and Grossmann \cite{Daubechies} discovered an
integral transformation similar to the well-known one of Bargmann and
Segal \cite{Bargmann}, but more directly adapted to quantization in that it
intertwined classical observables (i.e., functions on phase-space)
directly with the coherent-state transitions of the corresponding
quantized operators. They noted that the new transformation differed
from Bargmann's in two respects: the transformed operators acted on a
space with double the usual number of variables, and that some mixing
of the variables had occurred. We now show how these  phenomena may be
simply elucidated in terms of the symplectic groupoid
${\Bbb R}^{2n} \times \overline{\Bbb R}^{2n} \rightrightarrows
 {\Bbb R}^{2n}$.

   The idea is to pair the ``Moyal polarization'' $P$ of
Eq.~(\ref{polzns}) with a certain complex polarization $R$.
Specifically, write $z = x' + ix''$, $w = y' + iy''$, and take
\[
R = \text{span} \biggl\{ {\partial \over \partial \bar z},
                  {\partial \over \partial w} \biggr\}.
\]
Then $\overline P \cap R = 0$, and $K^R$ is spanned by
$dz \wedge d\bar w$. From Eq.~(\ref{Omeg}),
$\Omega = \frac i2 (dz \wedge d\bar z - dw \wedge d\bar w)$, and the
symplectic potential vanishing on $R$ is
\[
\Theta_R = - \case i2 (\bar z \,dz + w \,d\bar w).
\]

   Elements of $\Gamma_R L$ are of the form $h(z,\bar w)\,r_0$, where
$h$ is holomorphic in $(z,\bar w)$ and
$\nabla_X r_0 = - (i/\hslash)\,\Theta_R(X)r_0$. Thus $r_0 = \psi_0 t_0$
with $d\psi_0/\psi_0 = (i/\hslash)(\Theta_P - \Theta_R)$. It is
convenient to use the complex notationson the symplectic groupoid
$u = q_1 + ip_2$, $v = q_2 + ip_1$, and to write $d^2u = dq_1\,dp_2$,
etc. We thus get
\[
\psi_0 = C \exp\{ -(z\bar z + w\bar w + \bar uv - u\bar v)/4\hslash \}.
\]

   One finds that
$\langle \sqrt{dq_1\wedge dp_2}, \sqrt{dz\wedge d\bar w} \rangle = 1$,
so if $\gamma = h(z,\bar w) r_0 \otimes \sqrt{dz\wedge d\bar w}$, then
\begin{eqnarray*}
&&\langle\beta, \gamma\rangle = C \int \overline{g(u)} h(z,\bar w)
   \,e^{-(z\bar z + w\bar w + \bar uv - u\bar v)/4\hslash}
\\
&&= C \int \overline{g(u)} h(u + \case12 v,\bar u - \case12 \bar v)
   \,e^{-(2u\bar u + \bar uv - u\bar v + \case12 v\bar v)/4\hslash}
\\
&&= \langle g, Sh \rangle_{L^2({\Bbb R}^2)},
\end{eqnarray*}
with
\begin{eqnarray*}
Sh(u) &=& C \int h(u + \case12 v,\bar u - \case12 \bar v)
 \,e^{-(2u\bar u + \bar uv - u\bar v + \case12v\bar v)/4\hslash} \,d^2v
\\
&=& \int K(\bar z, w; u) h(z,\bar w)
     \,e^{-(z\bar z + w\bar w)/2\hslash} \,d^2z \,d^2w,
\end{eqnarray*}
where $K$ is computed from the reproducing kernel property of Gaussian
integrals:
\begin{eqnarray*}
K(\bar z, w; u)
&=& \frac{C}{(2\pi\hslash)^2} \int \exp\biggl(
    \frac{\bar z(u + \case12 v) + w(\bar u - \case12 \bar v)}{2\hslash}
\\
&&\quad - \frac{2u\bar u + \bar uv - u\bar v + \case12v\bar v}{4\hslash}
    \biggr) \,d^2v
\\
&=& \frac{2C}{\pi\hslash} \exp\bigl(
     (-2u\bar u + 2\bar zu + 2w\bar u - \bar zw)/2\hslash \bigr).
\end{eqnarray*}

   If $e_{\bar a,b}(z,\bar w) = \exp\{(\bar az + b\bar w)/2\hslash\}$
denote coherent-state vectors in $(z,\bar w)$-space, one checks that
$\|S e_{\bar a,b}\| = 2C(2\pi\hslash)^{3/2} \|e_{\bar a,b}\|$, so the
normalization $C = \case12 (2\pi\hslash)^{-3/2}$ makes $S$ unitary.
Moreover, $S^{-1}$ is given by the conjugate kernel:
\[
Q(z,\bar w; u) = \frac{2}{(2\pi\hslash)^{5/2}} \exp\bigl(
     (-2u\bar u + 2z\bar u + 2\bar wu - z\bar w)/2\hslash \bigr).
\]
Apart from Gaussian-integral conventions, this is precisely the kernel
of the Daubechies--Grossmann transformation which takes a Weyl symbol
$g$ to the coherent-state transition matrix:
\[
\langle w|Q_g|z \rangle = \int Q(z,\bar w; u) g(u) \, d^2u.
\]
Thus the symplectic groupoid picture shows that this arises from the
pairing of the polarizations $P$ and~$R$.

   The comparison with the double Bargmann transformation, explored in
\cite{Daubechies}, may now be clarified. The double Bargmann
transformation is obtained from the pairing of the polarizations $F$
and~$R$; the ``mixing'' of variables noted in \cite{Daubechies} comes
from the combinination of this pairing with that of Sec.~IV.

\section{Iteration of pairings}

In \cite{Phoebe} we proved, by a lengthy functional-analytic argument,
that the Weyl transform is of finite order. We now show that this comes
in fact from a simple identity among linear symplectomorphisms of the
groupoid.

   Let us write $q_1^{(0)} = x'$, $q_2^{(0)} = y'$, $p_1^{(0)} = x''$,
$p_2^{(0)} = -y''$, and considering the symplectic linear map $\Psi$
given by:
\begin{eqnarray}
&& q_1^{(1)} = {q_1^{(0)} + q_2^{(0)} \over \sqrt 2},\
   q_2^{(1)} = {p_1^{(0)} - p_2^{(0)} \over \sqrt 2},
\nonumber\\
&& p_1^{(1)} = {p_1^{(0)} + p_2^{(0)} \over \sqrt 2},\
   p_2^{(1)} = {q_2^{(0)} - q_1^{(0)} \over \sqrt 2},
\label{qpxy-two}
\end{eqnarray}
which is related to Eq.~(\ref{qpxy}) by $p_2 \mapsto q_2$,
$q_2 \mapsto -p_2$ and a rescaling by $\sqrt 2$ factors. The pairing
of the polarizations
$F^{(j)} = \text{span} \{ \partial/\partial p_1^{(j)},
                          \partial/\partial p_2^{(j)} \}$ $(j = 0,1)$
yields the unitary transformation of operator kernels:
\begin{eqnarray*}
&& Wg(q_1^{(0)},q_2^{(0)})
\\
&&= \frac 1{2\pi\hslash} \int g\biggl(
      {q_1^{(0)} + q_2^{(0)} \over \sqrt 2}, t \biggr)
      \,e^{it(q_1^{(0)} - q_2^{(0)})/\sqrt 2\hslash} \,dt.
\end{eqnarray*}
which is essentially the Weyl transformation: compare
Eq.~(\ref{Wigner-tr}).

   After three iterations of (\ref{qpxy-two}), the variables decouple
in two pairs:
\begin{eqnarray*}
&& q_1^{(3)} = {q_1^{(0)} + p_1^{(0)} \over \sqrt 2},\
   p_1^{(3)} = {- q_1^{(0)} + p_1^{(0)} \over \sqrt 2},
\\
&& q_2^{(3)} = {q_2^{(0)} + p_2^{(0)} \over \sqrt 2},\
   p_2^{(3)} = {- q_2^{(0)} + p_2^{(0)} \over \sqrt 2}.
\end{eqnarray*}
and $\Psi^6$ becomes simply:
\[
q_j^{(6)} = p_j^{(0)}, \ p_j^{(6)} = - q_j^{(0)},  \qquad  (j = 1,2),
\]
which is a complex structure on ${\Bbb R}^4$.  The pairing of $F^{(0)}$
and  $F^{(6)}$ yields the (inverse) Fourier transformation in the
variables $(q_1^{(0)},q_2^{(0)})$.

   It is well known \cite{Folland} that the Fourier transformation on
$L^2({\Bbb R}^n)$ is the image, under the metaplectic representation
of the symplectic group $Sp(2n,{\Bbb R})$, of the complex structure
$q \mapsto p$, $p \mapsto -q$ acting on Darboux coordinates on
${\Bbb R}^{2n}$. Now the symplectic group acts transitively on the set
of real polarizations of ${\Bbb R}^{2n}$, and the unitary
representation of the symplectic group given by pairing real
polarizations is precisely the metaplectic representation. Thus the
result of \cite{Phoebe} is now seen to be the metaplectic image of the
elementary geometric fact that $\Psi^6$ is a complex structure on the
symplectic groupoid ${\Bbb R}^2 \times \overline{\Bbb R}^2$, and thus
$\Psi^{24}$ is the identity map.

\acknowledgments

Heartfelt thanks to Philippe Blanchard for his warm hospitality at the
BiBoS Research Centre of the University of Bielefeld, where this work was
begun. Support by the Deutsche Akademische Austauschdienst of a visit
there by JCV is gratefully acknowledged. Support was also given by the
Vicerrector{\'\i}a de Investigaci\'on of the University of Costa Rica.

\end{document}